\documentclass[twocolumn, twocolappendix, numberedappendix]{openjournal}

\usepackage{lipsum}

\usepackage{xcolor}
\usepackage{textgreek}
\usepackage[utf8]{inputenc}
\usepackage[english]{babel}

\usepackage{hyperref}
\hypersetup{
    unicode, 
    colorlinks=true,
    linkcolor=linkcolor,
    citecolor=linkcolor,
    filecolor=linkcolor,
    urlcolor=linkcolor,
}
\usepackage{color,colortbl}
\definecolor{linkcolor}{rgb}{0.0,0.3,0.5}
\usepackage{tensind}
\tensordelimiter{?}
\DeclareGraphicsExtensions{.bmp,.png,.jpg,.pdf}
\usepackage{verbatim}
\usepackage[normalem]{ulem}
\usepackage{soul}

\defcitealias{Bondi2024}{B24}

\defcitealias{Kondapally21}{K21}
\newcommand*{\arcsecf}{\hbox{$.\!\!^{\prime\prime}$}}
\usepackage{booktabs}
\usepackage{float}

\graphicspath{ {./figures/} }

\shorttitle{Host galaxies of LOFAR sources in the EDFN}
\shortauthors{Bisigello et al.}

\begin{document}
\title{Host galaxy identification of LOFAR sources in the Euclid Deep Field North}

\author{\vspace{-4em}Bisigello,L.$^{\ast,1,2,3}$, Giulietti, M.$^{1}$, Prandoni, I.$^{1}$, Bondi, M.$^{1}$, Bonato, M.$^{1}$, Magliocchetti, M.$^{4}$, Rottgering, H.J.A. $^{5}$, Morabito, L.K.$^{6,7}$, White, G.J.$^{8,9}$}

\affiliation{\vspace{-0.7em}}
\affiliation{$^{1}$ INAF, Istituto di Radioastronomia, Via Piero Gobetti 101, 40129 Bologna, Italy}
\affiliation{$^{2}$Dipartimento di Fisica e Astronomia "G. Galilei", Universit\`a di Padova, Via Marzolo 8, 35131 Padova, Italy}
\affiliation{$^{3}$INAF--Osservatorio Astronomico di Padova, Vicolo dell'Osservatorio 5, I-35122, Padova, Italy}
\affiliation{$^{4}$INAF-IAPS, Via Fosso del Cavaliere 100, 00133 Roma, Italy }
\affiliation{$^{5}$Leiden Observatory, Leiden University, PO Box 9513, NL-2300 RA Leiden, The Netherlands.}
\affiliation{$^{6}$Centre for Extragalactic Astronomy, Department of Physics, Durham University, South Road, Durham DH1 3LE, UK}
\affiliation{$^{7}$Institute for Computational Cosmology, Department of Physics, Durham University, South Road, Durham DH1 3LE, UK}
\affiliation{$^{8}$Department of Physics and Astronomy, The Open University, Walton Hall, Milton Keynes, MK7 6AA, UK}
\affiliation{$^{9}$RAL Space, STFC Rutherford Appleton Laboratory, Chilton, Didcot, Oxfordshire, OX11 0QX, UK}

\email[$^\ast$]{laura.bisigello@inaf.it}

\begin{abstract}
We present a catalogue of optical and near-infrared counterparts to radio sources detected in the Euclid Deep Field North (EDF-N) using observations from the LOw-Frequency ARray (LOFAR) High Band Antenna (HBA) at 144 MHz with $6\arcsec$ angular resolution. The catalogue covers a circular region of $10\,\rm deg^2$ and includes 23\,309 radio sources with a peak signal-to-noise ratio greater than 5. After masking regions close to stars and with unreliable photometry in the optical or near-infrared, the catalogue includes 19\,550 sources. To carry out a robust identification strategy, we combined the statistical power of the Likelihood Ratio (LR) method, including both colour and magnitude information, with targeted visual inspection. The resulting catalogue boasts a remarkable identification rate of 99.2\%, successfully matching 19\,401 out of 19\,550 radio sources with reliable optical and/or near-infrared counterparts. 
For 19\,391 of the matched sources, we successfully derived photometric redshift for the host galaxy by performing an SED fit using the available data in the optical, near-infrared, far-infrared, and radio. LOFAR sources within the catalogue exhibit a median redshift of 1.1, with some extending up to $z=6$. Around 7\% of the sample is detected only in infrared using IRAC and tends towards higher redshifts, with a median of $z=3.0$. This comprehensive catalogue serves as a valuable resource for future research, enabling detailed investigations into the properties and evolution of LOFAR-detected sources and their host galaxies.
\end{abstract}


\maketitle

\section{Introduction}

Understanding the physics driving galaxy formation and evolution requires a deep understanding of baryonic processes, particularly those that regulate star formation and black-hole accretion across cosmic time. The radio sky offers a unique perspective on these processes, providing a direct view of both the build-up of stars and the formation and growth of supermassive black holes (SMBHs).

Radio surveys are effective methods to identify active galactic nuclei (AGNs), particularly in the presence of radio jets. When the SMBH is accreting gas at a high rate, about 10\% of the rest-mass energy of the accreting material is emitted as radiation. The dominant energetic output of these ‘radiative’ or ‘quasar-like’ AGN is in the form of electromagnetic radiation \citep{Heckman2014}. However, some of them are visible in the radio, thanks to the presence of powerful twin radio jets (radio-loud quasars and so-called high excitation radio galaxies), even though the mechanical energy remains lower than the radiation energy. Recent works have shown that even radiatively efficient AGNs that do not have powerful radio emission, called ‘radio-quiet’ AGNs, frequently possess weak radio jets \citep{Gurkan2019,Jarvis2019,Macfarlane2021,Morabito2022,Morabito2025}. As the accretion rate of material onto a SMBH decreases, the accretion flow is suggested to transit to a geometrically thick and radiatively inefficient state \citep{Narayan1994,Narayan1995}, releasing energy primarily through the two-sided radio jets. These ‘jet-mode' or ‘radio-mode' AGNs are visible in the radio, but, given that the radiative luminosity of the AGN is very low, their identification is challenging at other wavelengths. 

The study of complete AGN samples is pivotal to understanding their impact on the evolution of their host galaxy. Indeed, AGN feedback (radiative or mechanical) is expected to have an impact on the star formation of the host galaxies, locally preventing the gas from cooling and forming stars \citep{DiMatteo2005,Croton2006,McNamara2007,Fabian2012,Zubovas2012,Kormendy2013,Heckman2014,Hardcastle2020}. However, the impact (or lack thereof) of these feedback processes on the evolution of the host galaxy is still highly controversial \citep[e.g.,][]{Harrison2017,Magliocchetti2022}, which makes the analysis of AGN and the co-evolution with their host galaxy crucial.  

At the same time, when AGNs are not present, radio emission from galaxies at frequencies below $\simeq 1\,GHz$ is dominated by synchrotron radiation originated from electrons accelerated by recent supernova explosions associated with massive young stars \citep{Condon1992}. Thus, in this case, radio frequencies are tracing the star formation rate (SFR), over timescales of 100\,Myr \citep{Kennicutt2012}, without being affected by dust obscuration. This is particularly important taking into account that dust-obscured galaxies dominate the cosmic star-formation rate density at least up to $z=5$ \citep[e.g.,][]{Zavala2017}, but some very obscured sources may be present even at $z>6$ \citep[e.g.,][]{Rodighiero2023,Bisigello2023b}. Moreover, some objects are so deeply embedded in dust that they are faint or completely undetected in the rest-frame optical, and visible only at infrared (IR) or radio bands \citep{Wang2019,Gruppioni2020,Talia2021,Enia2022,vanderVlugt2022,Behiri2023,Gentile2024}, making observations at these frequencies extremely valuable for having a complete census of cosmic star formation at different cosmic epochs.

Identifying optical counterparts to radio sources is essential for maximising the scientific potential of deep extragalactic radio surveys. Detailed photometry from optical counterparts can be used for spectral energy distribution (SED) fitting, facilitating source classification, (photometric) redshifts estimation, and determining key physical parameters of the host galaxies, such as stellar masses and SFR. However, constructing complete, optically identified radio catalogues has proven challenging due to the limitation of radio observations. Indeed, the limited angular resolution of radio images can result in a poor positional accuracy, while the high density of optical and near-IR sources results in multiple potential counterparts. Moreover, radio sources often have large and intricate structures like jets, lobes, and hotspots, making the identification of the host position even more challenging. Taking this into account, a simple nearest-neighbour search is not always reliable, while the visual inspection of all sources becomes less and less feasible with the increasing size of radio surveys.

This paper presents the optical and near-IR counterparts of radio sources in the Euclid Deep Field North (EDF-N) obtained with the LOw-Frequency ARray (LOFAR) High Band Antenna (HBA). These counterparts are identified using the Likelihood Ratio (LR) method \citep{deRuiter1977,Sutherland1992}, which is a commonly used statistical technique to identify real counterparts of sources detected in observations with different angular resolutions \citep[e.g.][]{Ciliegi2003,Smith2011,McAlpine2012,Fleuren2012}. In particular, we use the colour-based adaptation of the LR method, developed by \citet{Nisbet2018} and used for the first data releases (DR1) of the LOFAR wide \citep{Williams2019} and LOFAR Deep \citep[][hereafter K21]{Kondapally21} surveys. The paper is structured as follows. In Section \ref{sec:data} we present the radio catalogue in the EDF-N as well as the available multi-wavelength observations. We outline the procedure to identify the counterpart of each radio source in Section \ref{sec:counterparts}, while we present the SED fitting method and the derived photometric redshift of these sources in Section \ref{sec:prop}. We finally report our conclusive remarks in Section \ref{sec:summary}. In this paper, we consider a $\Lambda$CDM cosmology with $H_0=70\,{\rm km}\,{\rm s}^{-1}\,{\rm Mpc}^{-1} $, $\Omega_{\rm m}=0.27$, $\Omega_\Lambda=0.73$, a Chabrier initial mass function \citep[IMF,][]{Chabrier2003}, and the AB magnitude system \citep{Oke1983}.

\section{Data}\label{sec:data}
In this Section we report the necessary details of the LOFAR catalogue and the ancillary data considered in this work. A summary of the number of sources available and the area covered by each dataset is listed in Table \ref{tab:Nsource}.

\begin{table}[]
\caption{Number of sources and area coverage of the ancillary catalogues used in this work.}
    \centering
    \begin{tabular}{ccc}
    Catalogue & Number of sources & Area\\
    \hline
        UV& 4\,657\,111 &14$\deg^{2}$\\    
        Optical & 25\,445\,387 & 44$\deg^{2}$\\
        \textit{Spitzer}/IRAC & 1\,806\,142 & 11.54$\deg^{2}$\footnote{This is the area covered by the observations in the [3.6$\,\rm\mu m$] band.}\\
        \textit{Herschel}/$70\mu m$ & 7 & \footnote{Derived by searching for objects within 6\arcsec from LOFAR sources.} \\
        \textit{Herschel}/$100\mu m$ & 472 & $^{\rm b}$ \\
        \textit{Herschel}/$160\mu m$ & 263 & $^{\rm b}$ \\
        \textit{Herschel}/$250\mu m$ & 3756 & \footnote{Derived by searching for objects within 12\arcsec from LOFAR sources.} \\
        \textit{Herschel}/$350\mu m$ & 2065 & $^{\rm c}$ \\
        \textit{Herschel}/$500\mu m$ & 615 & $^{\rm c}$ \\
    \end{tabular}
     
    \vspace{10pt}
    \label{tab:Nsource}
\end{table}

\subsection{Radio}
In this paper, we make use of an updated version (see Sect. \ref{sec:visualcheck}) of the catalogue presented by \citet{Bondi2024} and derived from 72 h of observations of the EDF-N obtained with the LOFAR High Band Antenna (HBA) at 144 MHz. The EDF-N is the fourth of the LOFAR Deep Fields and the data presented by \citet{Bondi2024} have similar sensitivity to the ones of the three other deep fields (Lockman Hole, ELAIS-N1 and Bo\"{o}tes) presented as part of the LOFAR Deep DR1 \citep{Tasse2021,Sabater2021}. The EDF-N LOFAR images have an angular resolution of 6\arcsec and a central rms noise of 32$\rm \,\mu Jy\,beam^{-1}$. The catalogue was extracted using the Python Blob Detection and Source Finder \citep[\texttt{PyBDSF},][]{PyBDSF}, which is a tool designed to decompose the radio images in islands above a given signal-to-noise (S/N) threshold for then fitting the islands with different components as Gaussians, shapelets, or wavelets. In this way, the extracted sources can be classified as "single" ($\rm S_{CODE}=S$) if they correspond to a single Gaussian component, "multiple" ($\rm S_{CODE}=M$) if they are described by multiple Gaussian components, or "complex" ($\rm S_{CODE}=C$) for a single-Gaussian source in an island with other sources. The catalogue includes 23\,309 sources with a peak $S/N>5$ in a circular region of 10 $\rm deg^{2}$. The majority of these sources ($93\%$) are classified as single, while 7\% are classified as multiple and only $<1\%$ are classified as complex.

To support and improve the multi-wavelength identification process of the radio sources we also make use of International LOFAR Telescope (ILT) images obtained including the international LOFAR stations. This image has been produced following the methods discussed in \citet{Sweijen2022} and \citet{deJong2024} and combines 32 out of the 72 h available. The final images cover the inner $2.5\times 2.5$ $\rm deg^{2}$ region of the EDF-N with a central sensitivity of 36$\rm \,\mu Jy\,beam^{-1}$ and an angular resolution of 1.5\arcsec.

\subsection{Gaia}
To mask sources whose photometry could be contaminated by stars and to verify possible astrometry offsets in the different catalogues, we retrieved objects in the Gaia data release 3 \citep{Gaia2016,Gaia_dr3} catalogue inside the LOFAR pointing. In particular, we recovered the sky position and g-band magnitudes of objects that are not identified as QSO or galaxies and have a $S/N>3$ in the parallax estimate. The retrieved GAIA catalogue contains a robust sample of 39\,885 stars, ranging from $g=6.7$ mag to $g=20.9$ mag.

\subsection{Optical}
The Hawaii eROSITA Ecliptic Pole Survey Catalog \citep[HEROES][]{Taylor2023} is based on Subaru Hyper Suprime-Cam (HSC) imaging observations covering 44~deg$^{2}$ in the North Ecliptic Pole. Observations were carried out in six broad-band and two narrow-band filters, reaching 5$\sigma$ depths of $g=26.5$, $r=26.2$, $i=25.7$, $z=25.1$, $y=23.9$, $\rm NB816=24.4$, and $\rm NB921=24.4$ mag. From the public catalogue, derived considering the standard hscPipe pipeline \citep[for more details on the catalogue extraction we refer to][]{Taylor2023} and including 25\,445\,387 sources, we selected only objects that are not detected as blended composite objects. In this way we removed objects that are artificially bright and for which a carefully deblending analysis would be necessary. We did not remove stars at this stage, as this step is performed later using the Gaia identifications. We verified the astrometry precision by cross-matching the catalogue with Gaia sources and we verified that the biases are below 0\arcsecf01 in both right ascension and declination. The final HEROES catalogue includes around $7\cdot10^{6}$ objects in the same 10~$\rm deg^{2}$ occupied by the LOFAR catalogue and with $S/N\geq3$ in at least one filter. Among all the different aperture magnitudes already present in the catalogue, we adopted the Kron aperture, as it takes into account the shapes of sources. We correct for galactic extinction using the central position of the LOFAR pointing as a reference and using the reddening values by \citet{Schlafly2011}.

\subsection{UV}
The Cosmic Dawn Survey \citep{Moneti2022} covers all \textit{Euclid} deep fields, including the EDF-N, with the Infrared Array Camera (IRAC) on board of the \textit{Spitzer} Space Telescope and also collects and provides ancillary ground-based observations \citep{McPartland2024}. The Cosmic Dawn Survey catalogue was derived using as detection image a combined $g+r+i$ image and it includes 5\,286\,829 sources. 
In particular, the survey includes $u$ band observations and, among all the sources in the catalogue, 4\,657\,111 are inside the $u$ band coverage and 3\,428\,854 have $S/N>3$ in the $u$ band. This band is fundamental to discriminate between $z<2$ galaxies and objects at higher redshifts. We did not include other bands from the Cosmic Dawn Survey has they are already present in the HEROES catalogue or are derived separately (see Appendix \ref{sec:IRAC}). The $u$ band data, whose fluxes are taken from the Cosmic Dawn Catalogue described in \citet{Zalesky2024}, are a collection of multiple surveys, mainly the Deep Euclid u-band Survey \citep[DEUS; see][]{Sawicki2019} and Hawaii Twenty $deg^{2}$ Survey (H20), and reach an average $5\sigma$ depth of $u=26.4$ mag. 

Matches with the HEROES optical catalogue are performed with a $1\arcsec$ radius, which is larger that the typical positional errors in the considered bands. We also verified this matching radius looking at the number of matches at increasing radii. Overall, 4\,035\,457 HEROES sources are in the $u$ band coverage and, among these, 3\,095\,639 have $S/N>3$. We correct for galactic extinction as done for the optical data.

\subsection{Near-IR}
As previously mentioned, the Cosmic Dawn Survey includes IRAC observations. This near-IR (NIR) survey covers 11.54, 11.74, 0.61 and 0.62\,deg$^{2}$ in the four IRAC channels, which are [3.6$\,\rm\mu m$], [4.5$\,\rm\mu m$], [5.8$\,\rm\mu m$], [8.0$\,\rm\mu m$], with different integration times, going up to 23.4h. The Cosmic Dawn Survey catalogue uses as detection image a combination of optical bands ($g+r+i$), but previous studies \citep[e.g.,][]{Nisbet2018, Williams2019} have shown that radio sources are preferentially hosted by red sources, indicating that it is fundamental to include also sources only detected in the IRAC bands. In addition, the Cosmic Dawn Survey catalogue is limited to the two IRAC filters at shortest wavelengths. For these reasons, we did not use the Cosmic Dawn Survey catalogue directly, but we perform a new extraction using the \texttt{Photutils} python package \citep{photoutils} on the the public release of the images presented in \citet{Moneti2022}. 

As a first step, we removed the background from each filter image.  We exploited the background estimation function (\texttt{Background2D}) of \texttt{Photutils}, adopting a box size of $32\times32$ pixels and a median filter of $3\times3$ pixels to create a background image from each filter image. The background was removed from each map to create a background-subtracted map for each IRAC channel. Then, we co-added the [3.6$\,\rm\mu m$] and [4.5$\,\rm\mu m$] images, considering the respective error maps, and we used this image to derive the segmentation map. A region was selected if it had at least three pixels above the threshold of three times the local rms noise. We further de-blended each segmentation region, considering as separate source peaks objects with a contrast of $10^{-8}$, with the contrast being the fraction of the total flux in the local peak, and with at least two pixels above the flux threshold. The contrast value was chosen between 0 (every local peak is a separate source) and 1 (no de-blending). This setup was fine-tuned by visually checking a subsample of possibly blended sources and we verified that it does not over-separate sources (i.e., does not split a single extended source into multiple point-like ones).
The segmentation map was then applied to each of the four filters separately to derive the flux of each source inside an elliptical Kron aperture. In particular, we considered a scaling parameter of the un-scaled Kron radius of 1.8, a minimum value for the un-scaled Kron radius of 2.5 pixels and no minimum circular radius. 

The final catalogue includes 1\,806\,142 sources that have a $S/N\geq3$ in the [3.6$\,\rm\mu m$] or [4.5$\,\rm\mu m$] filters. In Appendix \ref{sec:IRAC} we report a comparison between the IRAC magnitudes in our catalogue and in the Cosmic Dawn Survey one, showing they are overall consistent.

\subsection{Far-IR}
Using the IPAC infrared science archive, we looked for sources covered by far-IR (FIR) observations using the \textit{Herschel} telescope. We considered the PACS point source catalogue \citep{Marton2017} at 70, 100, and 160 $\mu m$ and the SPIRE Point Source Catalogue \citep{Schulz2018} at 250, 350, and 500\,$\mu m$. We searched for sources detected with PACS within 6\arcsec\, from each radio source, while we considered a search radius of 12\arcsec\, for SPIRE-detected sources. We retrieved 7 objects observed at $70\,\mu m$, 472 sources observed at $100\,\mu m$, 263 objects observed at $160\,\mu m$, 3756 sources observed at $250\,\mu m$, 2065 objects observed at $350\,\mu m$, and 615 sources observed at $500\,\mu m$. Overall, around 20\% of the LOFAR sources have observations in at least one FIR filter. Given the low number density of radio and FIR sources and the poor angular resolution of these observations, a more complex method for matching the sources is not required. At the same time, given that these FIR observations are relatively shallow, we do not attempt to de-blend sources.
\section{Radio-optical counterparts} \label{sec:counterparts}
In this section, we describe the method we adopt to associate a multi-wavelength identification to radio sources, which is developed starting from the work by \citet{Williams2019} and \citetalias{Kondapally21}. The method first masks areas with optical-to-NIR photometry contaminated by bright stars, then cross-matches optical and NIR catalogues to create a multiwavelength source catalogue, which is then associated with radio sources using a combination of likelihood ratio (LR) and visual inspection. We will now give more details for each of these steps. 

\subsection{Star masking}
Bright stars create spikes in optical and NIR images, making photometry unreliable. Following the method illustrated in \citetalias[][]{Kondapally21} (see their Section 3.4.3), we divided the stars of the Gaia catalogue into magnitude bins of 0.5. We used the radius-sky density variation plot as a diagnostic to determine a masking radius, which has been validated through visual inspection.
Briefly, the radius was chosen looking at the sky density of the sources as a function of the radius from the star in order to remove both the empty region of detections around stars and the spurious sources produced by the stars' spikes. The final radii of the circular masks vary from $178\arcsec$ for stars with $g\leq8.50$\,mag to 2\arcsecf0 (i.e., equal to the FWHM of the [4.5$\,\rm\mu m$] PSF in the \textit{Spitzer} warm setup) for stars with $g>21.1$\,mag. We masked a total area of 1.856 deg$^{2}$. The same mask is applied to the optical, NIR and radio catalogues, removing 3754 radio sources out of 23\,309 (i.e., $16\%$). Other 6 radio sources are also masked out as they are outside the footprint of the NIR observations, leaving a final catalogue of 19\,550 sources.  

\subsection{Optical-NIR cross-match}
In this section, we discuss the match between the optical HEROES catalogue and the NIR \textit{Spitzer} one. We applied a matching radius of $1\arcsec$, corresponding to half the FWHM of the [3.6$\,\rm\mu m$] \textit{Spitzer} band, and we consider only the closest match. Limiting the catalogue to sources with $16 < [4.5\,\rm\mu m] < 19$ to avoid faint or saturated sources, the astrometric offset between the optical and NIR catalogues are $\Delta(RA)=-0\arcsec.001_{-0.119}^{+0.120}$ and $\Delta(DEC)=0\arcsec.001_{-0.049}^{+0.048}$. This confirms the absence of strong biases and reassures that the radius used for the matching is more than eight times larger than the scatter of the positional offsets. The final catalogue consists of 1\,610\,079 sources with optical-to-NIR data in the LOFAR area, of which 75\,890 (5\%) have at least one other possible counterpart within $1\arcsec$. We include a flag in the catalogue for all the IRAC sources with possible optical secondary counterparts. In addition to these, we also included 198\,255 IRAC sources without an optical counterpart and 3\,976\,651 optical sources without a NIR counterpart. 

\subsection{Improved LOFAR astrometry}\label{sec:offsets}
\begin{figure}
    \centering
    \includegraphics[width=\linewidth]{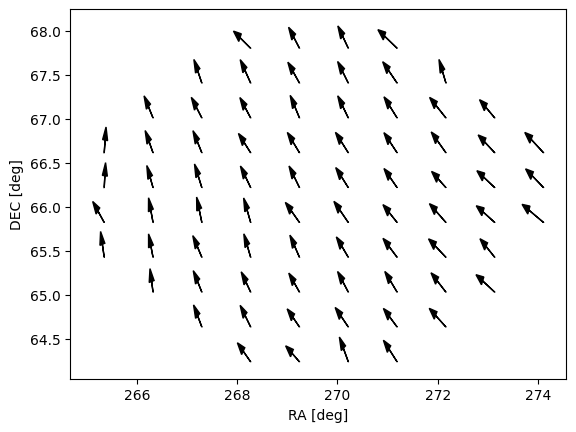}
    \caption{Spatially resolved offsets between LOFAR sources and the closest optical-NIR objects in the chosen sub-regions (see text for details). The length of each arrow is multiplied by a factor of 1000 for visual purposes.}
    \label{fig:offsets}
\end{figure}

Before proceeding with the derivation of the likelihood ratio, we estimated any relative astrometry offset between the LOFAR and optical-NIR data. This is done for the catalogue derived from the $6\arcsec$ image, as higher angular resolution images are available only for a portion of the field. To do so, we find the closest match to each radio source in the optical-NIR catalogue, removing objects in the masked regions. We then divide the LOFAR pointing in a grid of 10$\times$10 subregions, each of $0.97\times0.39\,\rm deg^{2}$. Relative offsets are then derived in each subregion. The median offset (radio-optical) corresponds to $\Delta(RA)=-0\arcsecf47^{+0.19}_{-0.26}$ and $\Delta(DEC)=0\arcsecf47^{+0.04}_{-0.03}$. When a subregion includes less than ten objects, we apply the median offset of the entire pointing.
These offsets, which are applied to the LOFAR coordinates and included in the final catalogue, are shown in Fig. \ref{fig:offsets}.

The subregions we consider cover an area similar to the facets used by \citep{Bondi2024} to perform the direction dependent calibration, but they are not coincident with them. To verify that this different does not impact the astrometric correction, we check that differences between adjacents subregions are smooth. Indeed, the median values of the differences in the correction is $\Delta[\Delta (RA)]=0\arcsecf04$ and $\Delta[\Delta(DEC)]=-0.0007\arcsec$, with maximum values of $0\arcsecf5$ in right ascension and $0\arcsecf2$ in declination, for regions generally close to the edge of the field. In addition, we also check how the astrometric corrections change by dividing the field in 9$\times$9 and 11$\times$11 subregions. The 1, 2 and 3 $\sigma$ values of the distribution of the difference in right ascensions are $0\arcsecf1$, $0\arcsecf2$, and $0\arcsecf5$, respectively. The 1, 2 and 3 $\sigma$ values of the distribution of the difference in declination are instead even smaller, that is $0\arcsecf03$, $0\arcsecf07$, and $0\arcsecf16$, respectively.

\subsection{Deriving the likelihood ratio}
The LR is a statistical method to identify counterparts of sources in images with poor angular resolution \citep{deRuiter1977,Sutherland1992}, which is often the case at radio or IR wavelengths \citep[e.g.,][]{Smith2011,Fleuren2012,McAlpine2012}. We followed the iterative approach reported in the \citetalias{Kondapally21} for the other LOFAR DR1 Deep fields and we refer to their paper and \citet{Williams2019} for more details. 

Briefly, the LR is defined as the ratio between the probability that a galaxy with a given set of properties is the true counterpart and the probability that it is an unrelated background object. Given that previous studies \citep[e.g.,][]{Nisbet2018, Williams2019} have shown that radio sources are more likely hosted by red sources, we use both the magnitude ($m$) and the colour ($c$) to derive the LR, as follows:
\begin{equation}
    LR(m,c)=\frac{q(m,c)f(r)}{n(m,c)}.
\end{equation}
In particular, $q(m,c)$ indicates the prior probability that a source with a specific magnitude $m$ and colour $c$ is the true counterpart of a LOFAR source. $f(r)$ is instead the probability distribution of the offset between the LOFAR source and the possible counterpart and takes into account their positional uncertainties. Finally, $n(m,c)$ is the sky density of all galaxies of magnitude $m$ and colour $c$. Given that radio sources are more probably hosted by red sources, it is preferable to consider the colour derived from an optical and a near-IR band. In this work we chose the $i$ and the [4.5$\,\rm\mu m$] bands for consistency with other Deep DR1 LOFAR fields. Following \citetalias{Kondapally21}, we derived the LR for all possible counterparts within $10\arcsec$ from each LOFAR source. An iterative process allows to derive the LR for each possible counterpart and the LR threshold above which a counterpart is considered valid. 

In particular, a first iteration is performed considering separately the $i$ and the [4.5$\,\rm\mu m$] magnitudes and using sources with $\rm S_{CODE}=S$, with major axis sizes smaller than $10\arcsec$ and with $S/N>3$ in the $i$ and the [4.5$\,\rm\mu m$] bands, separately. This allow us to obtain a first estimate of the LR threshold above which we consider a counterpart correct. The fractions of radio sources that have a reliable counterpart $Q_{0}$ after these first LR estimations are $Q_{0,i}=0.92$ and $Q_{0,[4.5\,\rm\mu m]}=0.98$. These are in lines with the values found by \citetalias{Kondapally21} for the other Deep DR1 LOFAR fields (i.e. from 0.75 to 0.95). 

\begin{figure}
    \centering
    \includegraphics[width=\columnwidth,keepaspectratio]{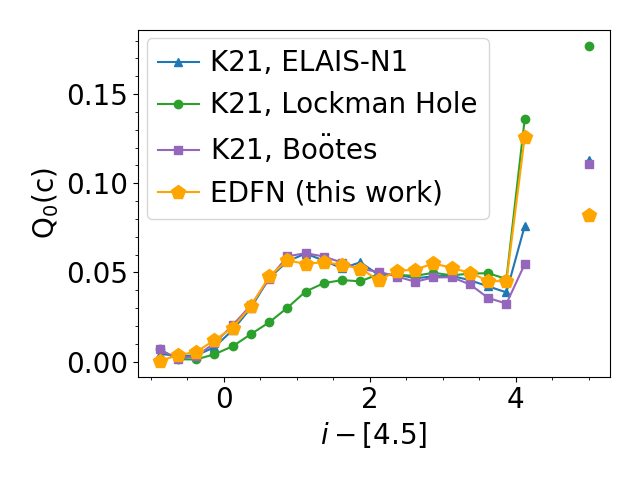}
    \caption{Fraction of genuine counterparts $Q_{0}(c)$ as a function of $i-[4.5\,\rm\mu m]$ colour for the EDF-N (this work) and other LOFAR deep fields \citepalias{Kondapally21}. Optical-only sources, IRAC-only sources and sources with $S/N<3$ in both the $i$ and [4.5$\,\rm\mu m$] bands, but detected in other optical or NIR filters, are arbitrarily set at $i-[4.5\,\rm\mu m]=-1$, $i-[4.5\,\rm\mu m]=4$, and $i-[4.5\,\rm\mu m]=5$, respectively.}
    \label{fig:Q0c}
\end{figure}

These first estimates are then used to derive the colour-magnitude LR and the corresponding threshold, considering again only radio sources with $\rm S_{CODE}=S$ and major axis sizes smaller than $10\arcsec$, but without imposing any $S/N$ limit in the optical-NIR catalogues. We derived a new estimate for the LR threshold, considering that $Q_{0}=\sum_{c} Q_{0}(c)$ with $Q_{0}(c)$ the fraction of radio sources that have a reliable counterpart of colour $c$, and a new set of accepted counterparts. The entire second step is then iterated until no changes are found in the matches, which in our case happened at the seventh iteration. As visible in Fig. \ref{fig:Q0c}, the final values for $Q_{0}(c)$ are in general agreement with those found by \citetalias{Kondapally21} in the Deep DR1 LOFAR fields. This plot confirms that radio sources are preferentially hosted by red sources, that is $i-[4.5\,\rm\mu m]>0$. Overall, we found a $Q_0=0.996$ corresponding to a LR threshold of 0.0079. 

\subsection{Validate the LR with visual inspection}\label{sec:visualcheck}
\begin{figure}
    \centering
    \includegraphics[width=\linewidth]{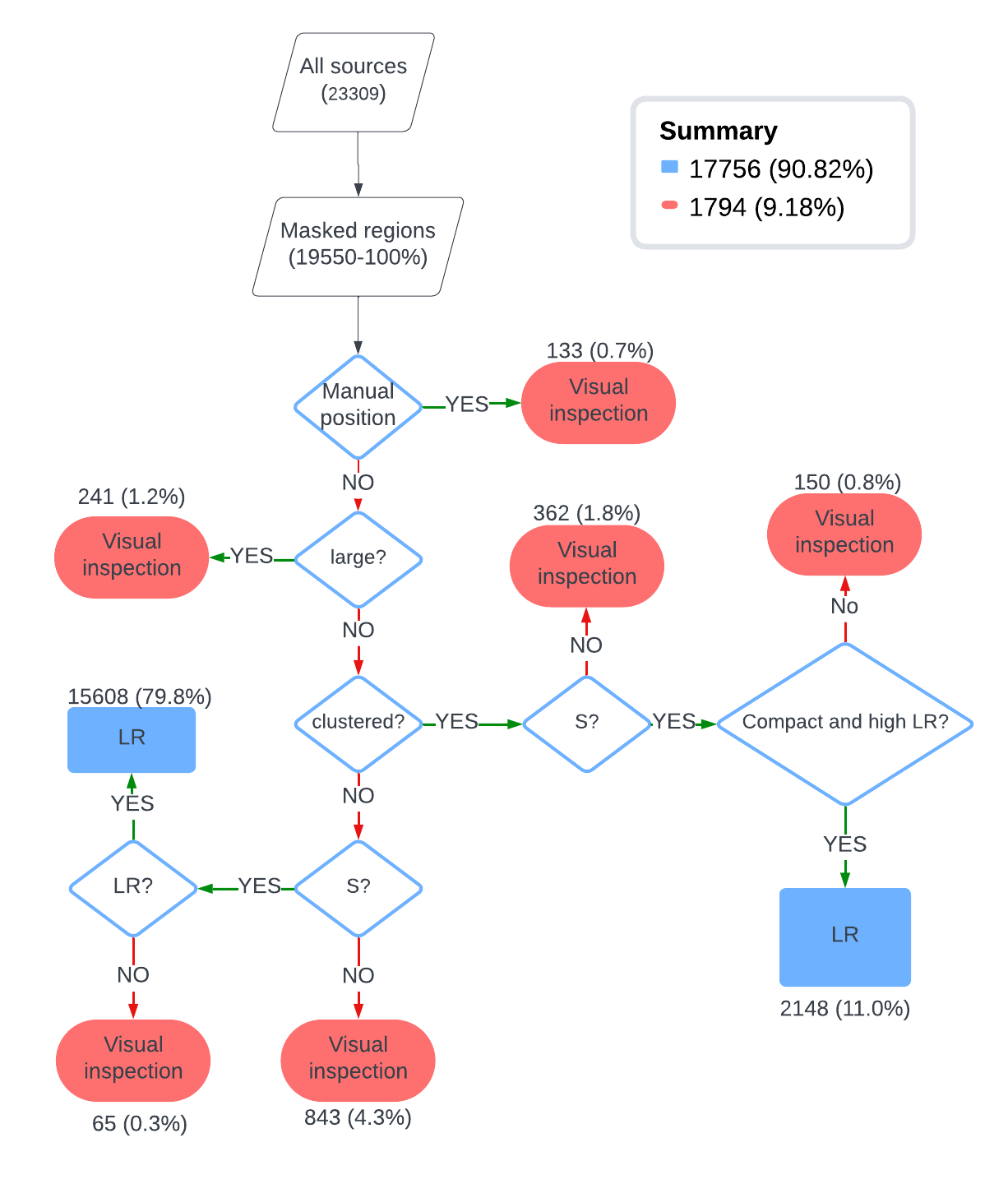}
    \caption{Flowchart showing the process followed to decide if a source requires a visual inspection (red ovals) or if we can rely directly on the LR (cyan squares). These are the decision knots: i) Manual position: the position was not derived with \texttt{PyBDSF}, but set manually; ii) Large: major axis larger than 15$\arcsec$; iii) Clustered: $4^{th}$ neighbour with 45$\arcsec$; iv) S: "single" morphology defined as a single Gaussian using \texttt{PyBDSF}; v) Compact and high LR: major axis smaller than 10$\arcsec$ and $LR>10\,LR_{threshold}$; vi) LR: $LR>LR_{threshold}$. LB{Numbers refer to the updated version of the catalogue by \citet{Bondi2024}.}}
    \label{fig:LRtree}
\end{figure}

In this work, we used a decision tree similar to the one presented by \citetalias{Kondapally21} to decide when we could rely directly on the LR to identify the multi-wavelength counterpart of a radio source and when, instead, we needed to validate the LR match with a visual inspection. The decision tree is summarised in Fig. \ref{fig:LRtree}. The main difference with respect to \citetalias{Kondapally21} is that all dubious cases are visualised by a group of experts, without requiring the assistance of citizen science. This was possible because of the reduced number of sources that required this check.

\begin{figure}
    \centering
    \includegraphics[width=\linewidth]{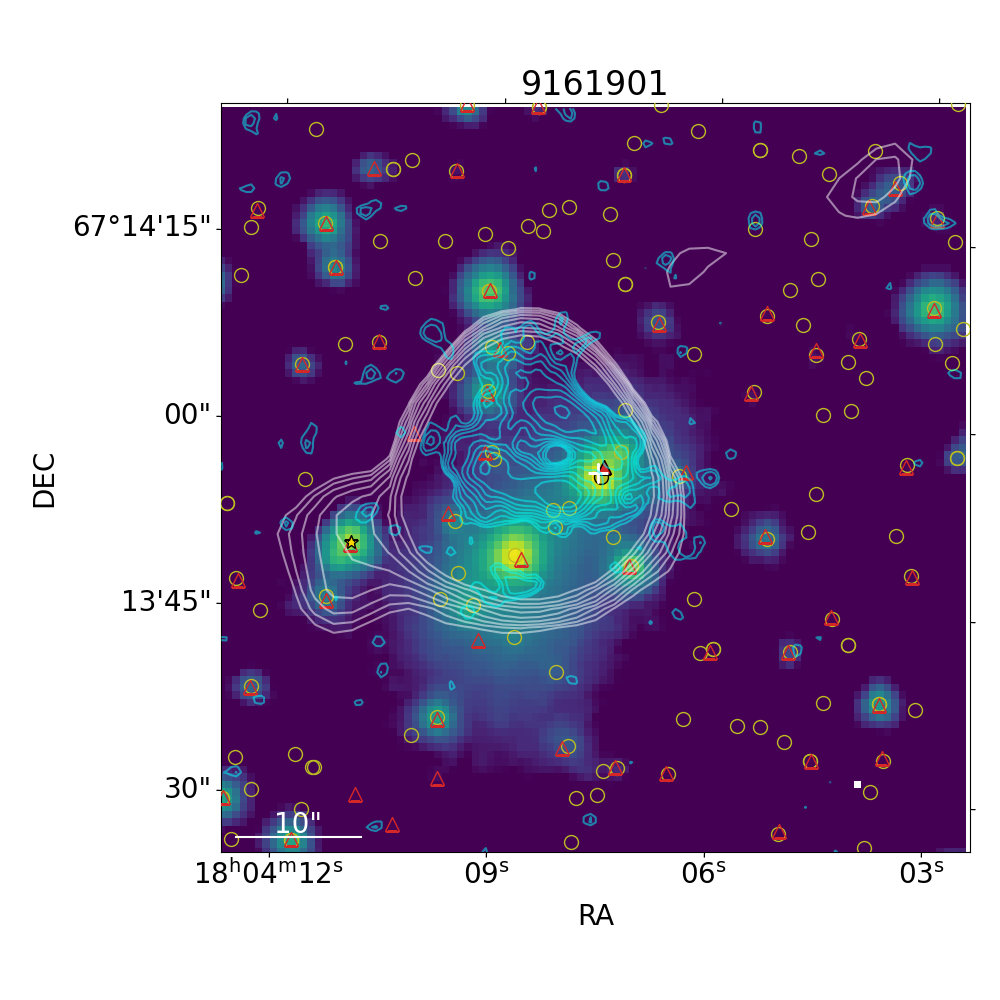}
    \caption{Example of a radio source for which the position was updated thanks to the $1\arcsecf5$ resolution LOFAR image. On the background, we show the [4.5$\,\rm\mu m$] image with superimposed the radio contours, from $3\times$ to $20\times$ the noise in logarithmic scale, at 6\arcsec (white contours) and 1\arcsecf5 (cyan contours) resolution. The old position of the radio source is indicated with a magenta cross, while the final position is shown with a white cross. We do not report the positions of other LOFAR sources. Red empty triangles show the positions of all IRAC sources, while yellow empty circles refer to the optical sources. The filled symbols show the position of the recognised optical and NIR counterparts. Yellow stars show the position of known GAIA stars.}
    \label{fig:pos_change}
\end{figure}

In particular, we visually inspected:
\begin{itemize}
    \item sources that are neither clustered, nor large, nor have a morphology that is described as "single" by \texttt{PyBDSF} ($\rm S_{CODE}=S$). The environment is defined as having the fourth neighbour within $45\arcsec$. Large sources have \texttt{PyBDSF} major axis larger than $15\arcsec$. These are 843 sources, that is $4.3\%$ of the sample (here and later we refer to the sample after applying the masking). 

    \item sources that are not large, but are in a clustered environment and do not have a "single" morphology. These sources, which are 362 ($1.8\%$ of the sample), could potentially be part of large complex systems, making it difficult to pinpoint the host.

    \item large sources with \texttt{PyBDSF} major axis larger than $15\arcsec$, which are typically complex or have poor positional accuracy. These are 241 sources corresponding to 1.2\% of the sample.

    \item sources that are clustered, have a "single" morphology, appear to be clearly resolved ($r>10\arcsec$), even if the do not fall into the aforementioned "large source", having $r<15\arcsec$, or their LR is not very robust (i.e., $LR<10\,LR_{threshold}$). This corresponds to 150 objects, which is $0.8\%$ of the sample.
        
    \item sources for which \texttt{PyBDSF} did not find the position of the radio core which, therefore, required a manual identification, potentially increasing systematic offsets. These are 133 sources corresponding to 0.7\% of the sample.
    
    \item sources that are neither clustered, nor large, have a "single" morphology, but the LR is below the defined threshold. These are 65 objects, corresponding to 0.3$\%$ of the overall sample.
\end{itemize}
There is no overlap between the mentioned sub-samples of sources requiring visual inspections. Overall, we visually inspect 1794 sources, equal to $9.18\%$ of the entire sample. Each source was visually inspected by at least three people to verify if the LOFAR source was correctly identified and if the optical-NIR counterpart was plausible. Given the iterative nature of this procedure, precise statistics on this analysis can not be given.
When available, to support our visual inspection, we also exploited a higher-spatial resolution LOFAR image, derived at 1\arcsecf5 angular resolution. In particular, we used these higher resolution image to better estimate the position of the radio core in all cases where the 6\arcsec resolution image did not allow for a clear identification and when we had blended radio sources. For the subsample of objects with available 1\arcsecf5 angular resolution, the original source parameters presented by \citet{Bondi2024} have been re-evaluated, taking  into account also the IRAC images, as \citet{Bondi2024} considered Wide-field Infrared Survey Explorer \citep[WISE;][]{Wright2010} data with lower angular resolution. This re-evaluation can also affect the number of LOFAR sources, as in some cases the components of a multiple source can get split into individual  sources or vice versa. As a result the original 23\,333 sources reported by \citet{Bondi2024}, get reduced to 23\,309, as reported in this paper. We made the updated version of the \citet{Bondi2024} LOFAR catalogue available\footnote{Add link}. Figure \ref{fig:pos_change} shows an example of a galaxy for which the sky position was corrected thanks to the high-resolution data. 

The final LOFAR multi-wavelength catalogue contains 19\,401 objects out of 19\,550, corresponding to 99.2\% of the original catalogue, with a reliable optical and/or NIR counterpart, with only 149 objects having a non-reliable counterpart. As summarised in Table \ref{tab:N_id}, the majority of the sources have both optical and NIR counterparts, with only a handful of them having only optical fluxes and 7\% having only IRAC fluxes. A description of the released catalogue is presented in Appendix \ref{sec:cat}. 

In Figure \ref{fig:Sdist} we show the number of radio sources with identifications as a function of radio flux density and the identification method used (LR or visual inspection). As visible, the LR is mainly used for faint sources, which not surprising given they are generally unresolved, while the visual inspection is dominant at $S>10\,\rm mJy$. A similar split is also present in the other LOFAR DR1 Deep fields \citepalias{Kondapally21}. 

\begin{figure}
    \centering
    \includegraphics[width=\linewidth]{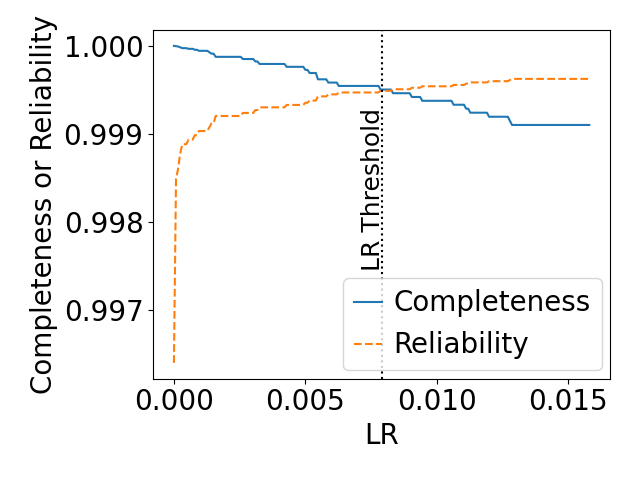}
    \caption{Reliability (orange dashed line) and completeness (blue solid line) derived for the subsample of objects that rely directly on the LR. The vertical dotted line is the LR threshold, which corresponds to the intersection point between completeness and reliability.}
    \label{fig:RC}
\end{figure}

The reliability of matches that rely directly on the LR is expected to be very high, above 99.9\% for LR larger than the threshold, as shown in Figure \ref{fig:RC}. The reliability is however expected to be worse for complex or faint systems, as it is discussed later for IRAC-only sources.

\begin{figure}
    \centering
    \includegraphics[width=\linewidth]{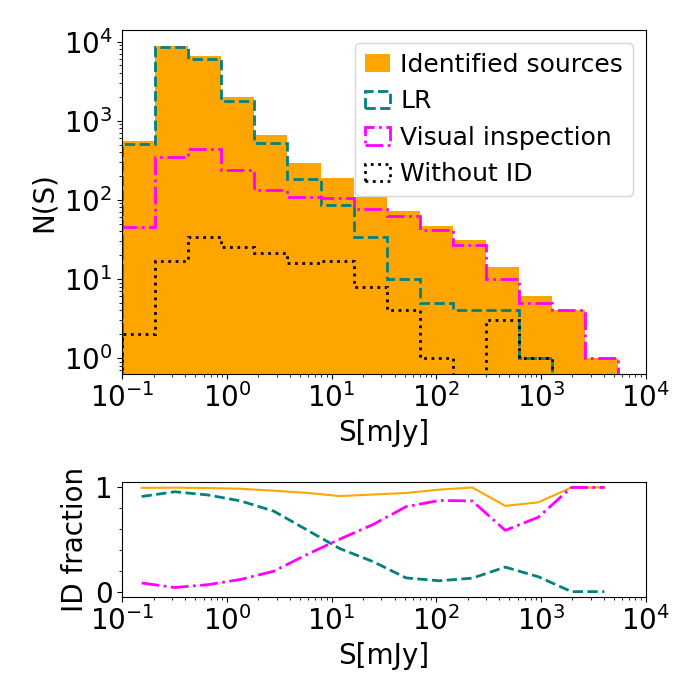}
    \caption{Top: Number of radio sources with valid identifications as a function of radio flux density and the identification method used (LR or visual inspection). For comparison, we also show the flux density distribution of the sources without a multi-wavelength identification (black dotted line). Bottom: identification fraction (ID) as a function of the flux density and the identification method. The fraction is derived with respect to the total number of sources, with or without identification, after applying masking. }
    \label{fig:Sdist}
\end{figure}

\begin{table}[]
    \caption{Summary of the number of sources in the LOFAR catalogue with and without counterparts.}
    \centering
    \begin{tabular}{c|c}
        Sample & Number of sources \\
        \hline
        Original \citep{Bondi2024} & 23\,333 \\
        Updated\footnote{The parameters of some LOFAR sources have been updated considering radio images at 1\arcsecf5 angular resolution and IRAC images.} & 23\,309 \\
        After masking & 19\,550 \\
        With identifications & 19\,401\\
        Optical+NIR identifications & 18\,032 \\
        Optical-only & 6 \\
        IRAC-only & 1363 \\
        Not identified & 149\\
    \end{tabular}
    \label{tab:N_id}
\end{table}

\section{Properties of host galaxies}\label{sec:prop}
\subsection{Magnitude distributions}
\begin{figure*}
    \centering
    \includegraphics[width=\linewidth]{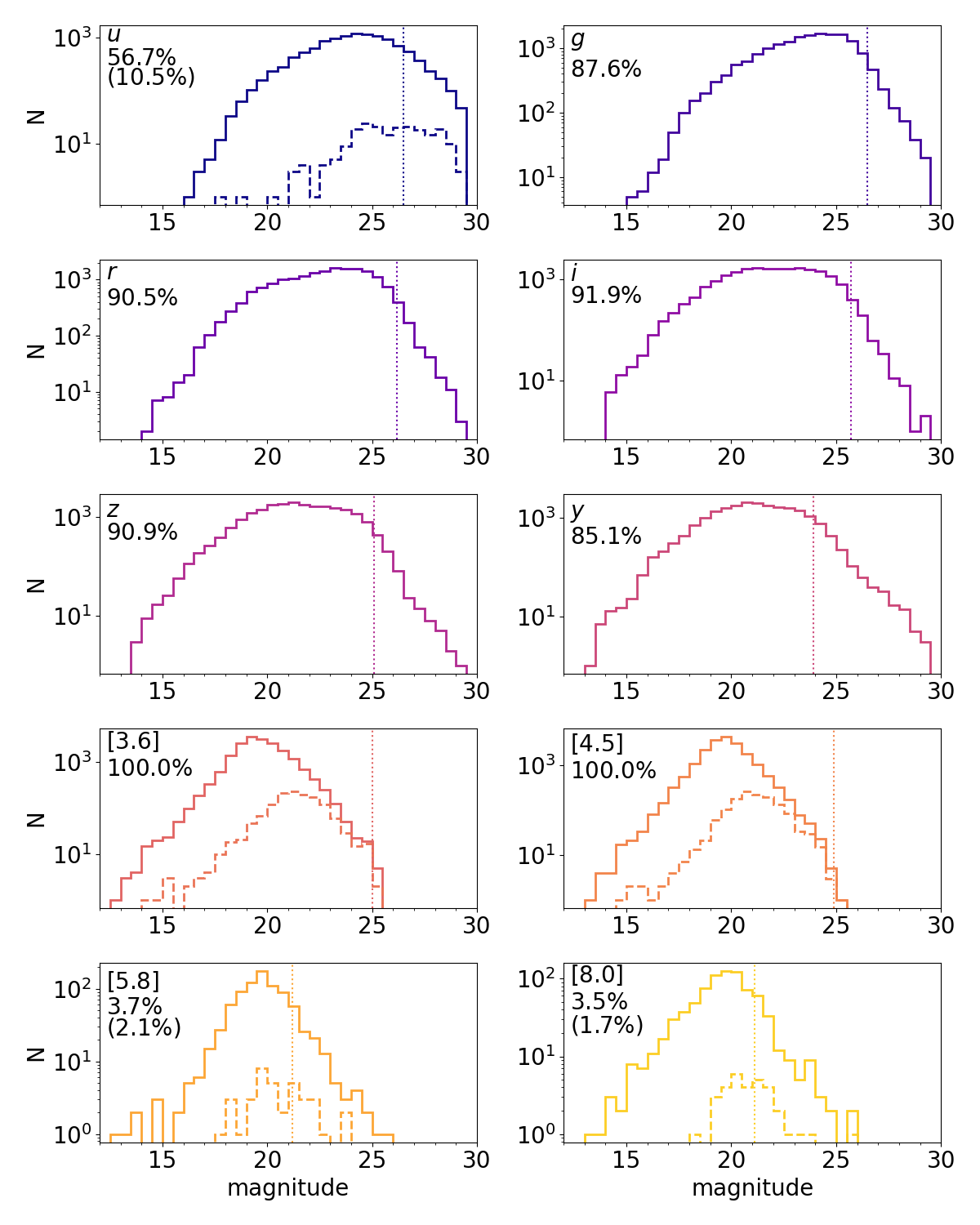}
    \caption{Magnitude distribution of the counterparts of the LOFAR sources in the different UV, optical, and NIR bands. Dashed histograms show the distributions of the IRAC-only sources. In each panel we also report the fraction of counterparts with $S/N>3$ in that particular filter. The fraction between brackets refer to the IRAC-only sources. We remind the reader that the [5.8$\,\rm\mu m$] and [8.0$\,\rm\mu m$] observations are available only for a small part of the EDF-N, that is $\sim0.6\,\rm deg^2$. The vertical dotted lines indicate the $5\sigma$ depth of each filter.}
    \label{fig:mag}
\end{figure*}

In Figure \ref{fig:mag} we report the magnitude distributions of the counterparts of the LOFAR sources. We can see that, as a result of the red colours of the host galaxies, 85\% to 91\% of them are detected ($S/N>3$) in the optical bands, 57\% are detected in the $u$ band, while all of them are detected in the IRAC [3.6$\,\rm\mu m$] and [4.5$\,\rm\mu m$] bands. The two IRAC bands at 5.8 and 8.0 $\mu m$ cover a small fraction of the LOFAR pointing, resulting in less than $4\%$ of the counterparts being detected in these bands. The magnitude distributions show peaks between the 20th and 25th magnitudes, with fainter peak magnitudes reached at longer wavelengths, and a decrease at fainter magnitudes. These peaks are driven by the depth of the LOFAR observations, as the $5\sigma$ of each band are deeper, as shown by the vertical dotted lines in the figure. Overall, the NIR observations generally match the LOFAR depth, while deeper optical observations would be necessary.

\subsection{Radio sources with IRAC-only identifications}
As previously mentioned, 7\% of the LOFAR objects, corresponding to 1363 sources, lack a detection in the optical filters in the HEROES catalogue and have instead a counterpart in the IRAC bands. Among these IRAC-only sources, 218 have a flux in the $u$ band from the Cosmic Dawn Catalogue, even if only 143 have a $S/N>3$ in the same band. Therefore, a small fraction of IRAC-only sources are probably at the limit depth of optical observations and a different extraction technique may derive some estimate for optical fluxes. In Figure \ref{fig:mag} we show the magnitude distributions of the IRAC-only sources in the $u$ band and the IRAC filters, where it is possible to compare with the magnitude distribution of the overall LOFAR sample. From this Figure it is evident that the IRAC-only sources are in general among the faintest sources in the LOFAR matched catalogue.

\begin{figure}
    \centering
    \includegraphics[width=\linewidth]{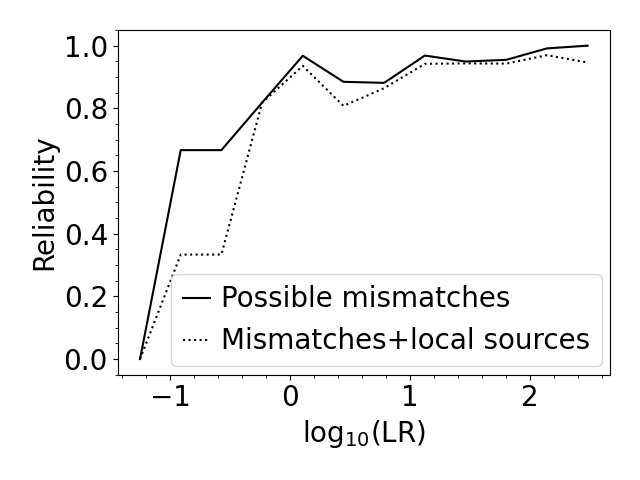}
    \caption{Reliability derived by visually inspecting IRAC-only sources. The solid line indicates the reliability obtained as one minus the fraction of sources with a possible wrong match, due to blending, crowding, or complex radio morphology. The dotted line which are mainly found by considering also local galaxies that have been wrongly over separated.}
    \label{fig:R_Io}
\end{figure}

We visually inspected all IRAC-only sources, as they are the faintest and possibly less reliable objects. Indeed, 66 out of 1363 could possibly be wrong identifications, due to blending issues in the IRAC bands or to a particular crowded field. In addition, we identified 28 objects for which the optical fluxes were wrong because a local extended galaxy was split into multiple source. This splitting is due to the setup of the extraction of the optical catalogue, which favours small compact sources over large ones. We use these mismatches to compute the reliability of the counterpart for the IRAC-only sources, shown in Figure \ref{fig:R_Io}, as a proxy for the reliability of complex sources. A reader particularly interested in a very pure sample can use this estimate to cut the LR at a higher value. We however remind that the matches of less problematic objects, which are the majority of sources in the LOFAR catalogue, are expected to have a very high reliability even at lower LR (Fig. \ref{fig:RC}).

\subsection{Radio sources without an identification}
\begin{figure}
    \centering
    \includegraphics[width=\linewidth]{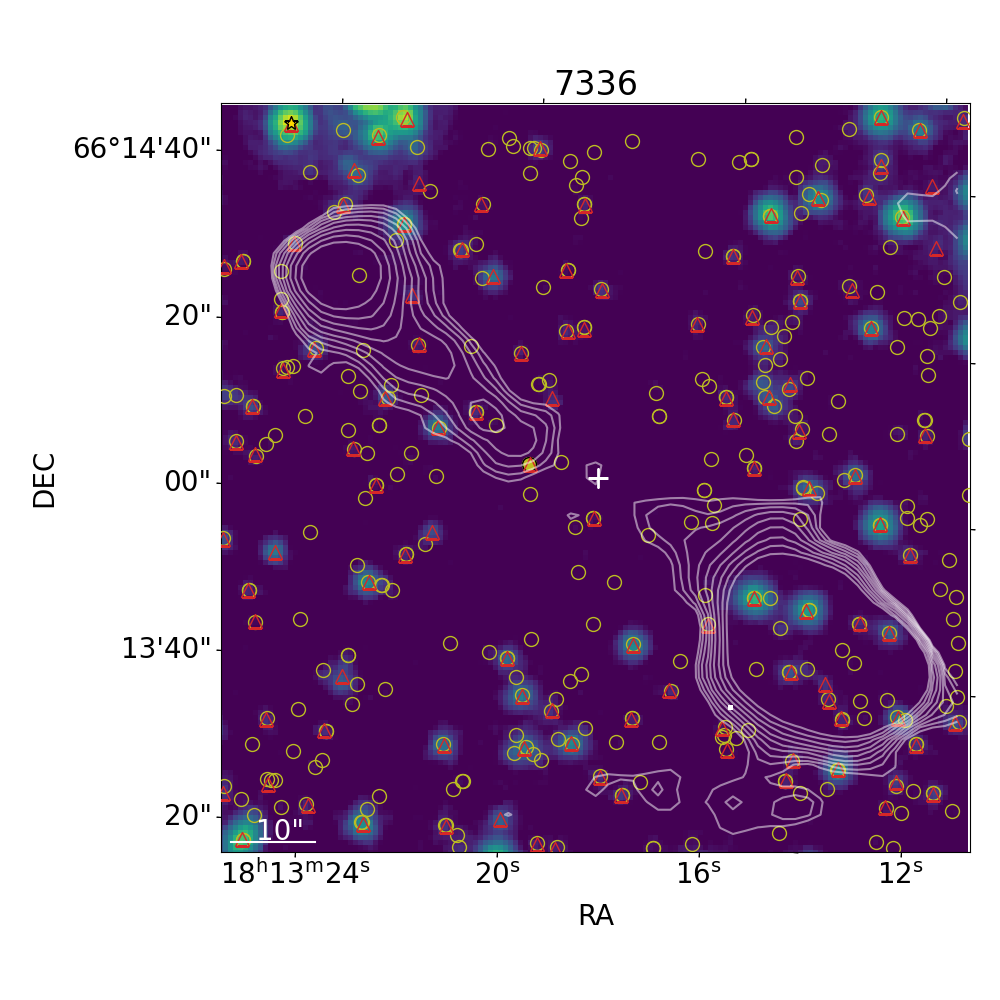}
    \caption{An example of a LOFAR source without a reliable optical or NIR counterpart. On the background, we show the [4.5$\,\rm\mu m$] image with superimposed the radio contours at $6\arcsec$ resolution, from $3\times$ to $20\times$ the noise in logarithmic scale. The position of the radio source is indicated with a white cross. Red empty triangles show the positions of all IRAC sources, while yellow empty circles refer to the optical sources. The filled symbols show the position of the recognised optical and NIR counterparts, which is unreliable being below the LR threshold. Yellow stars show the position of known GAIA stars. The LOFAR source has a total flux of $S=64.2\pm0.4\,\rm mJy$. }
    \label{fig:noID}
\end{figure}

A minority of sources in the LOFAR catalogues, that is 149 objects, remains without reliable optical-NIR counterparts. They are all sources that went through the visual inspection. The majority of them (76/149) were visually inspected because their likelihood ratio was below the threshold. Of the remaining, 32 do not have a single morphology based on \texttt{PyBDSF}, 23 are clustered, 21 are large, 11 have a manual position, and 9 are compact and isolated but with a low likelihood ratio. The radio fluxes of these sources is shown in Fig. \ref{fig:Sdist}. The median radio fluxes is 1.79\,mJy, slightly above the median value of the entire sample of visually inspected objects, that is 0.86\,mJy. They therefore represent a mixture of sources for which it was difficult to identify the counterparts because of a complex morphology or because the area was very crowded, or because they are sources for which current optical and NIR ancillary data are too shallow. An iconic example for the latter case is presented in Figure \ref{fig:noID}, where we show a LOFAR source with two powerful jets, but no clear counterpart present in the proximity of the centre of the jets.

\subsection{SED fitting}
To derive the physical properties of the identified sources to perform some preliminary test, we performed SED fitting of optical-to-radio data with the \texttt{CIGALE} code \citep{Buat2018}. We leave to a future paper the comparison with other codes. The complete list of free parameters considered in \texttt{CIGALE} is reported in Table \ref{tab:cigalegrid}. As input to the fit, we consider all the ancillary photometry from optical to FIR, when available, and the LOFAR fluxes, given that the latest \texttt{CIGALE} version includes a modelling of radio emission separated into two possible components: one associated with star formation and one associated with a radio AGN \citep{Yang2022}. The first is derived considering the IR-radio correlation parameter \citep[e.g][]{Helou1985,Condon1992,Yun2001,Delhaize2017,Delvecchio2021}, which we kept fixed to the default value of $q_{ir,sf} = 2.58$, while the latter is parametrised with the radio-loudness parameter, defined as $R_{AGN}=L_{\nu,\rm 5GHz}/L_{\nu,2500\AA}$, where $L_{\nu,2500\AA}$ and  $L_{\nu, \rm 5GHz}$ are the monochromatic AGN luminosities at $2500\AA$ and 5GHz. It is necessary to consider that only 20\% of the sources have a detection in the FIR, so the division between star formation and AGN contribution in the radio may be uncertain. To be conservative, we added in quadrature a relative error of 0.1 to the uncertainties of the fluxes.

LOFAR galaxies in this work have on average 8.5 filters with $S/N\geq3$, indicating a good coverage from optical to radio. However, 7\% of the matched sources are detected only in the IRAC bands and, therefore, have a limited photometric coverage, with 3.4 filters on average with $S/N\geq3$. These are mainly the LOFAR and the IRAC [3.6$\,\rm\mu m$] and [4.5$\,\rm\mu m$] photometric points. Therefore, physical properties derived for these sources need to be considered with caution and additional NIR observations, like the one that will be available with the Euclid mission \citep{Mellier2024}, will be fundamental to better characterize these sources.

\begin{figure*}
    \centering
    \includegraphics[width=0.48\linewidth]{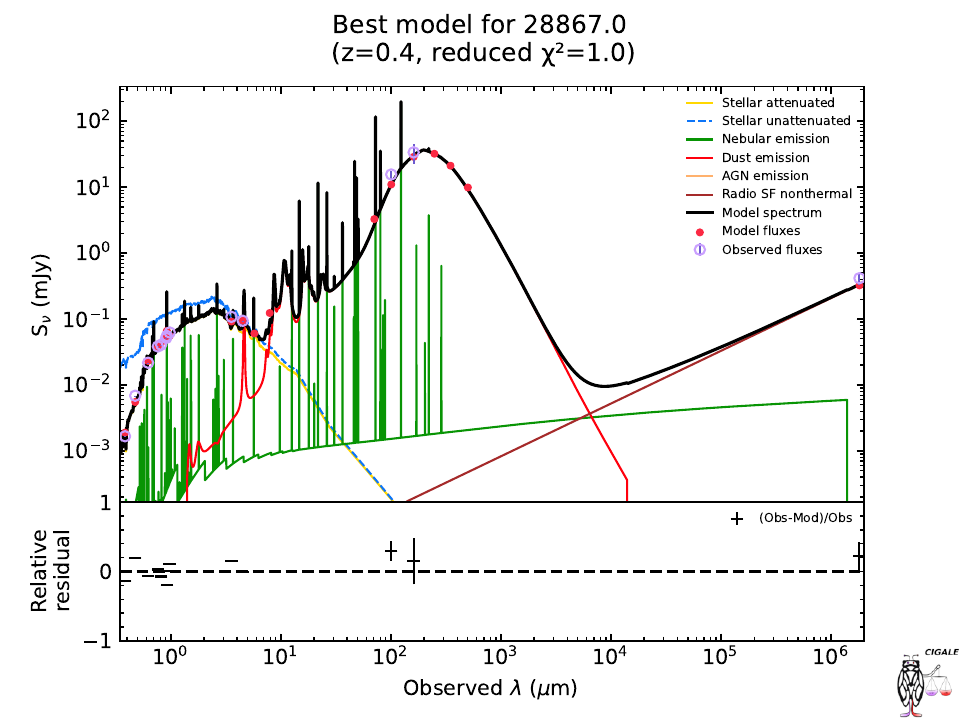}
    \includegraphics[width=0.48\linewidth]{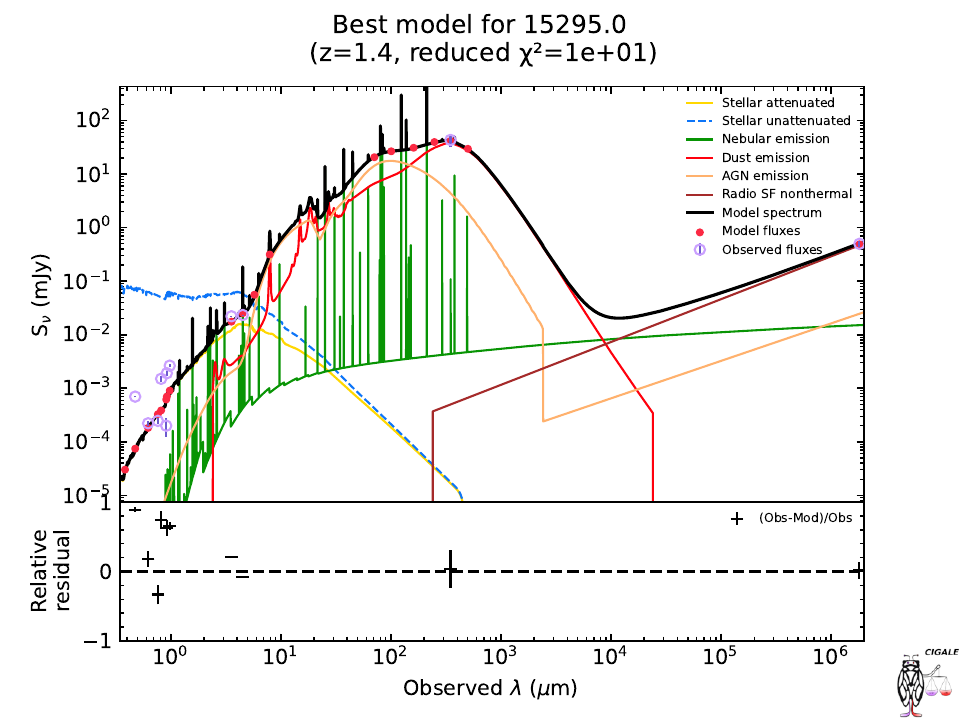}
    \includegraphics[width=0.48\linewidth]{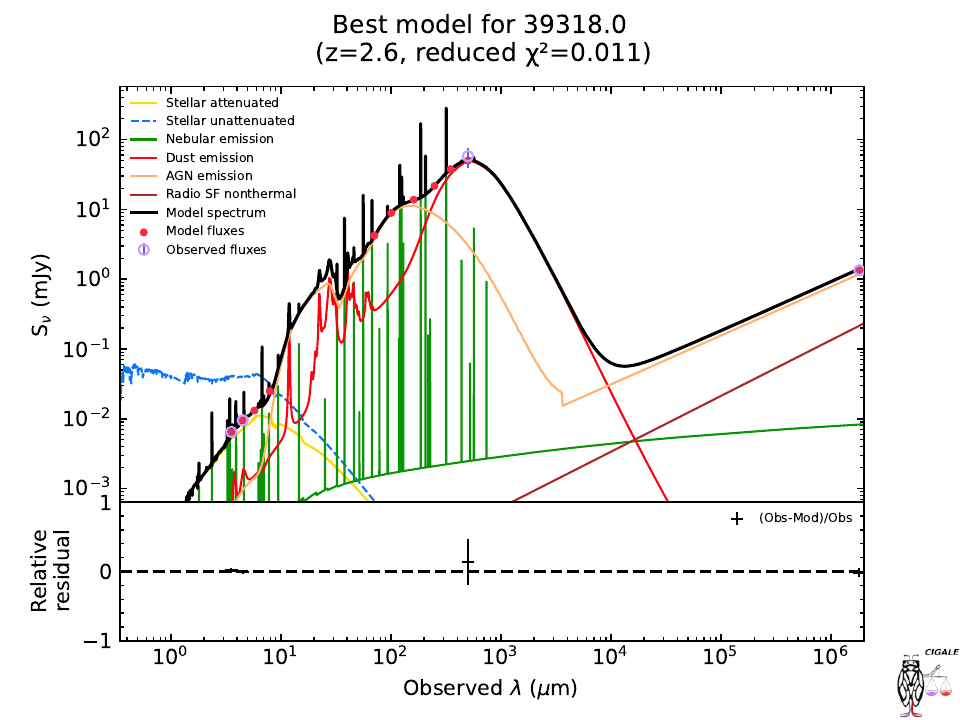}
    \caption{Three examples of SED fit. In the top part of each panel we show the observed fluxes (purple open circles), the expected fluxes from the model (red filled circles) and the different SED components (coloured lines). In the bottom part of each panel we show the residuals. We report an object with good fit and photometric coverage (top left), an object with many photometric points but a poor SED fit (top right), an IRAC-only source with an unconstrained fit of an IRAC-only source (bottom). }
    \label{fig:SED}
\end{figure*}

Overall, 19396 out of 19401 LOFAR sources with an optical and/or NIR counterpart have a good SED fitting, resulting in a $\chi_{red}^{2}<10$. All optical-only and IRAC-only sources also have a $\chi_{red}^{2}<10$. However, we note that in both cases these sources are affected by an artificially low $\chi_{red}^{2}$ due to the reduced number of bands with a detection. We report some examples of good, bad and unconstrained SED fits in Figure \ref{fig:SED}.

\begin{table*}
    \centering
    \caption{Grid utilised for the free parameters of our \texttt{CIGALE} SED-fitting run. All the fixed parameters are not showed and their values coincide with the default ones assigned by \texttt{CIGALE}.}
    \begin{tabular}{p{5.5cm} p{4.5cm} p{6cm}} 
        \textbf{\texttt{CIGALE} fit parameters} & \textbf{Grid values} & \textbf{Description} \\
        \cmidrule(lr){1-3}
        \multicolumn{3}{l}{\textit{Delayed SFH} [\texttt{sfhdelayedbq} module]} \\
        $\tau_{\rm main}$ & 200, 300, 500, 700, 1000, 1500 & e-folding time of the main stellar population model in Myr \\
        Age & 200, 300, 500, 700, 1000, 1500 & Age of the main stellar population in the galaxy in Myr \\
        $Age_{burst}$ & 10, 30, 100 & Age of the burst/quench episode, in Myr. \\
        $f_{burst}$ & 0.1, 1, 10 & Ratio of the SFR after/before $Age_{burst}$ \\
        \cmidrule(lr){1-3}
        \multicolumn{3}{l}{\textit{Dust attenuation component} [\texttt{dustatt\_modified\_CF00} module]\footnote{\citet{CF00}}} \\
        $A_{\rm V, ISM}$ & 0.01, 0.25, 0.5, 1.0, 3. & V-band attenuation in the interstellar medium\\
        \cmidrule(lr){1-3}
        \multicolumn{3}{l}{\textit{AGN component} [\texttt{fritz2006} module]\footnote{\citet{Fritz2006}}} \\
        $\tau$ & 0.6, 1.0, 3.0 & Optical depth at 9.7$\mu m$ \\
        $\Psi$ & 0.001, 89.99 & Angle between the equatorial axis and line of sight \\
        $\text{f}_{\rm AGN}$ & 0., 0.25, 0.5 & AGN fraction with respect to the total dust luminosity \\
        \cmidrule(lr){1-3}
        \cmidrule(lr){1-3}
        \multicolumn{3}{l}{\textit{Radio emission} [\texttt{radio} module]} \\
        $R_{AGN}$ & 0.01, 1., 100., 10000. & The radio-loudness parameter for AGN, defined as $R_{AGN}=L_{\nu,\rm 5GHz}/L_{\nu,2500\AA}$, where $L_{\nu,2500\AA}$ is the AGN $2500\AA$ intrinsic disk luminosity measured at viewing angle=$30^{\circ}$. \\
        \cmidrule(lr){1-3}
        $z$ & 0. to 6. with a step of 0.1 & redshift \\
        \label{tab:cigalegrid}
    \end{tabular} \\ 
\end{table*}

\subsection{Photometric redshifts}
In this paper we discuss the photometric redshifts, while we leave the analysis of the other physical properties to a future work. The redshift estimations are reported in the released catalogue (see Appendix \ref{sec:cat}).

\begin{figure}
    \centering
    \includegraphics[width=\linewidth]{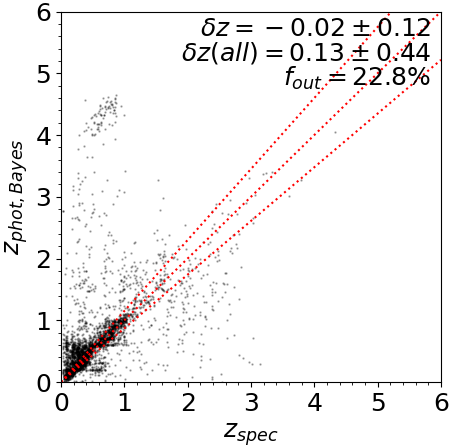}
    \includegraphics[width=\linewidth]{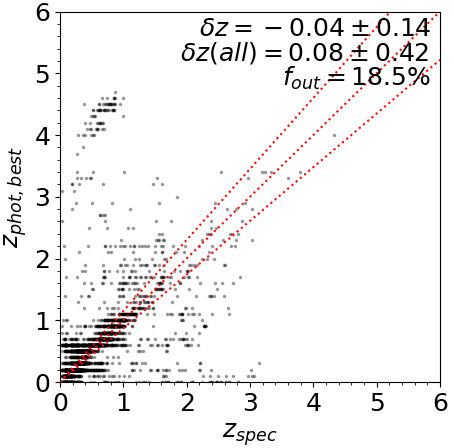}
    \caption{Comparison between the spectroscopic redshfits from DESI and the photometric redshifts derived using \texttt{CIGALE}. Bayesian-like photometric redshifts are at the top and best photometric redshifts are at the bottom. In the upper right of each panel, we report the mean and standard deviation, without and with the outliers defined as objects with $|\delta z|=|(z_{phot}-z_{spec})/(1+z_{spec})|>0.15$, together with the outlier fraction. }
    \label{fig:zspec}
\end{figure}

To validate the photometric redshifts derived with the SED fitting procedure, we compare them with the spectroscopic redshifts available from the Dark Energy Spectroscopic Instrument \citep[DESI;][]{DESI2024}. The matching is performed using the position of the optical or near-IR counterpart and adopting a matching radius of 0\arcsecf5. We obtained 4034 matches, corresponding to 21\% of the LOFAR sources with reliable optical-to-NIR counterparts. The comparison, shown in Fig. \ref{fig:zspec}, is limited to galaxies and QSO with good-quality spectroscopic redshifts (i.e. $\rm ZWARN=0$) and a $\chi_{red}^{2}<10$ from the SED fitting. As for the SED fitting, we consider two different estimates included in \texttt{CIGALE}. The first one corresponds to the value of the best SED template, which is the one with the minimum $\chi^{2}$, while the second photometric redshift estimate is derived with a Bayesian-like approach \citep[see][for more details]{Noll2009}. 

\begin{figure}
    \centering
    \includegraphics[width=\linewidth]{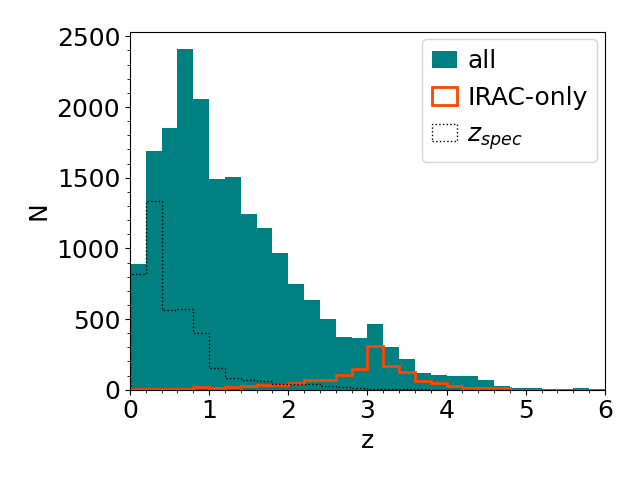}
    \caption{Redshift distribution of LOFAR sources with optical and/or NIR counterparts and a good SED fit (i.e. $\chi_{red}^{2}<10$). The distribution is shown for the complete sample (green filled histogram), for galaxies without an optical counterpart (orange solid histogram), and for galaxies with spectroscopic redshifts from DESI (black dotted histogram).}
    \label{fig:zdist}
\end{figure}

As visible in Figure \ref{fig:zspec}, when considering the Bayesian-like approach, photometric redshifts include $22.8\%$ outliers, defined as objects with $|\delta z|=|(z_{phot}-z_{spec})/(1+z_{spec})|>0.15$, which are mainly found among galaxies with $z_{spec}<1$, where however the outlier definition turns out to be more stringent. For some of the outliers, the discrepancy decreases if we consider the $z_{phot}$ associated with the best SED model instead of the Bayesian estimate, reducing their fraction to 18.5\%. The redshift uncertainties associated with outliers are large, indicating that these are difficult sources with multiple photometric redshift solutions. In addition, some of the optical/NIR largest sources, which are at low redshifts, may suffer from fragmentation, as the catalogue creation in the optical is optimized for compact sources. We verified that the difference between photometric and spectroscopic redshift does not correlate with the number of bands available. However, only three IRAC-only sources have a spectroscopic redshift. Indeed, as visible in Fig. \ref{fig:zdist}, the redshift distribution of galaxies with spectroscopic redshift is biased towards low redshift values. In the future, the inclusion of Euclid photometry and informative priors can help limiting the main degeneracies and improve the redshift recovery. For example, \citet{Enia2025} has reported an outlier fraction of 7\% using $ugriz$, Euclid and two IRAC bands ([3.6],[4.5]).

Taking into account the uncertainties previously mentioned, in Figure \ref{fig:zdist} we report the overall redshift distribution. LOFAR sources show a median $z=1.1$ and extend up to $z=6$, which is the maximum redshift considered in the fit. Galaxies presenting a counterpart in IRAC, but not in the optical, are skewed towards higher redshifts, with a median value of $z=3.0$. This is in line with other observations of optically faint or optically-dark objects \citep[e.g.,][]{Wang2019}. However, as already noted, it is necessary to take these results with caution, given that, by construction, the SED fitting of these sources is based on a limited number of detections (3.4 filters with $S/N>3$, on average). As an additional estimate of the redshift accuracy, we have considered sources for which the best and Bayesian redshift of \texttt{CIGALE} agree, taking the normalised difference of 0.15 as threshold. We have consistent redshift estimates for 89\% of the overall sample, while this fraction is reduced to 39\% for the IRAC-only subsample. 

\begin{figure}
    \centering
    \includegraphics[width=\linewidth]{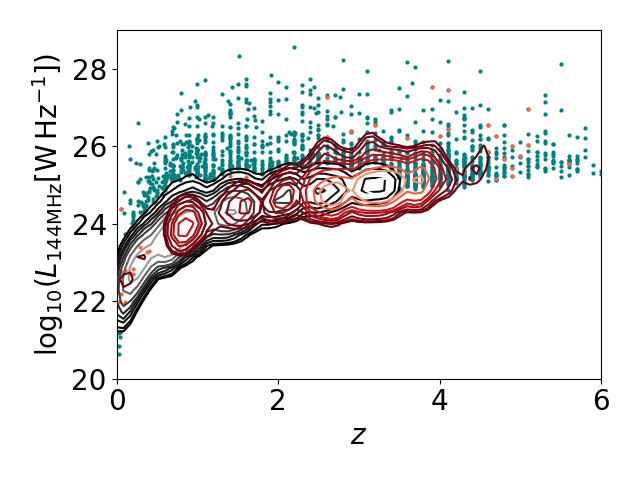}
    \caption{Radio luminosity as a function of redshift, spectroscopic or photometric, depending on the availability. We considered only objects for which the two \texttt{CIGALE} photometric redshift estimates agree (normalised difference less 0.15). Contour lines are logarithmically spaced from 10\% to 90\% of the distribution and correspond to the entire sample (black lines) or the IRAC-only sources (red lines). The remaining 10\% of the sample is shown with coloured points. Red points indicate IRAC-only sources.}
    \label{fig:Lradio}
\end{figure}

We also report in Figure \ref{fig:Lradio} the 144MHz radio luminosity of LOFAR sources, considering the spectroscopic redshift when available, and the photometric redshift otherwise. We show this luminosity only for sources with a consistent best and Bayesian photometric redshifts. To convert from observed to rest-frame frequencies we considered a spectral index of $\alpha=-0.7$, with no variation with redshift \citep{CalistroRivera2017}. We considered a single spectral index for all sources, as we leave the classification of the objects in star-forming galaxies or AGN to a future work. Anyway, considering a spectral slope of $\alpha=-0.664\pm0.011$ or $\alpha=-0.729 \pm 0.010$, as derived by \citet{CalistroRivera2017} for LOFAR-selected AGN and star-forming galaxies respectively, would produce a change of less than $6\%$ on the radio luminosity.

\begin{figure}
    \centering
    \includegraphics[width=\linewidth]{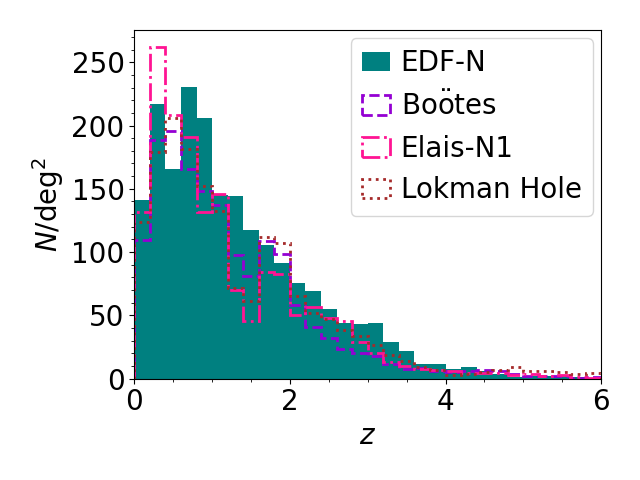}
    \caption{Normalised redshift distribution of sources with fluxes $S>0.3\,\rm mJy$ in the four LOFAR deep fields.}
    \label{fig:z_LDF}
\end{figure}

Finally, in Figure \ref{fig:z_LDF} we compare the normalised redshift distribution of LOFAR galaxies in the EDF-N with the one of other LOFAR deep fields \citep{Duncan2021}. Given the different depths of the LOFAR deep fields, we consider for this comparison only sources with fluxes $S>0.3\,\rm mJy$. Overall, the redshift distributions are quite similar, with the EDF-N having a median redshift $z=1.03$, while the other three fields have $z=0.92$ (Bo\"otes), $z=0.82$ (ELAIS-N1), and $z=0.97$ (Lokman Hole).

\section{Conclusions}\label{sec:summary}
In this paper, we have presented a catalogue of optical and NIR counterparts to radio sources in the Euclid Deep Field North observed with the LOFAR High Band Antenna at $6\arcsec$ spatial resolution over an area of $10\,deg^{2}$. A high identification rate of 99.2\% is achieved through a multi-faceted approach. We used the Likelihood Ratio (LR) method, adapted to incorporate both colour (i.e, $i-[4.5\,\rm\mu m]$) and magnitude information,  following the approach presented by \citet{Nisbet2018} and employed in prior LOFAR surveys \citep{Williams2019,Kondapally21}. This statistical method provides a superb performance in identifying counterparts for isolated, compact radio sources. We complement this analysis with visual inspection for sources with complex morphology, large sizes, or in particularly crowded areas. When possible, in the visual inspection we made use of LOFAR observations at higher angular resolution, reaching 1\arcsecf5, to improve the positional accuracy and the counterpart identification. The resulting LOFAR catalogue contains 19\,401 objects with reliable optical and/or NIR counterparts out of the initial 19\,550, demonstrating the effectiveness of the identification strategy. 

In the paper, we further characterise the identified sources by deriving their photometric redshift via an SED fitting analysis. In the fit we consider, when available, data in the optical, NIR, FIR and radio. The LOFAR sources exhibit a median redshift of 1.1, extending up to $z = 6$. Sources detected solely in the infrared bands through IRAC are shifted towards higher redshifts, with a median of $z = 3.0$, but their photometric redshifts need to be considered with caution, given the few filters with $S/N>3$ available. 

Overall, this study provides a comprehensive catalogue of optical and NIR counterparts for LOFAR radio sources in the EDF-N, laying the foundation for further investigations into the properties and evolution of these sources and of their host galaxies. The availability of Euclid data in the upcoming future will be crucial in refining the properties of these sources, especially of those only detected in the NIR. At the same time, it will be fundamental to complement these radio data with FIR observations, like the ones that could be achieved with the Probe far-Infrared Mission for Astrophysics (PRIMA, P.I. J. Glenn), in order to have a better characterization of both the star-forming and the AGN component \citep{Bisigello2024}.

\section*{Acknowledgments}

LB acknowledges support from INAF under the Large Grant 2022 funding scheme (project "MeerKAT and LOFAR Team up: a Unique Radio Window on Galaxy/AGN co-Evolution"). The research activities described in this paper were carried out with contribution of the Next Generation EU funds within the National Recovery and Resilience Plan (PNRR), Mission 4 - Education and Research, Component 2 - From Research to Business (M4C2), Investment Line 3.1 - Strengthening and creation of Research Infrastructures, Project IR0000034 – “STILES - Strengthening the Italian Leadership in ELT and SKA”.  LKM is grateful for support from a UKRI FLF [MR/Y020405/1] and STFC grant [ST/V002406/1].
This work has made use of data from the European Space Agency (ESA) mission
{\it Gaia} (\url{https://www.cosmos.esa.int/gaia}), processed by the {\it Gaia}
Data Processing and Analysis Consortium (DPAC,
\url{https://www.cosmos.esa.int/web/gaia/dpac/consortium}). Funding for the DPAC
has been provided by national institutions, in particular the institutions
participating in the {\it Gaia} Multilateral Agreement. This research made use of Photutils, an Astropy package for detection and photometry of astronomical sources \citep{photoutils}.

\bibliographystyle{aa} 

\bibliography{oja_template}

\begin{appendix}

\section{Comparison with Cosmic Dawn Survey catalogue}\label{sec:IRAC}
In this Section we show some comparison between the IRAC magnitudes derived in this work, using the $\rm [3.6\,\rm\mu m]+[4.5\,\rm\mu m]$ image to detected sources, with the IRAC magnitudes of the Cosmic Dawn Survey catalogue, derived using the $r+i+z$ image to detected sources. Our catalogue includes 1\,529\,783 objects with $S/N>3$ in the $[3.6\,\rm\mu m]$ or $[4.5\,\rm\mu m]$ band and outside the masked regions. The uncertainties in the IRAC bands of the Cosmic Dawn Survey catalogue are underestimated, as pointed out by \citet{Zalesky2024}, so we perform a magnitude cut in the $[3.6\,\rm\mu m]$ or $[4.5\,\rm\mu m]$, corresponding to a $S/N>3$, resulting in 2\,934\,425 sources. Considering a matching radius of $1\arcsec$ we found 1\,325\,106 objects in common. We notice that we have been more conservative while masking areas around stars, so many sources of the Cosmic Dawn Survey catalogue are located in areas we masked. On one side, using an optical prior they extracted sources at fainter magnitudes. On the other side, our catalogue includes sources without an optical counterpart.

\begin{figure}
    \centering
    \includegraphics[width=\linewidth]{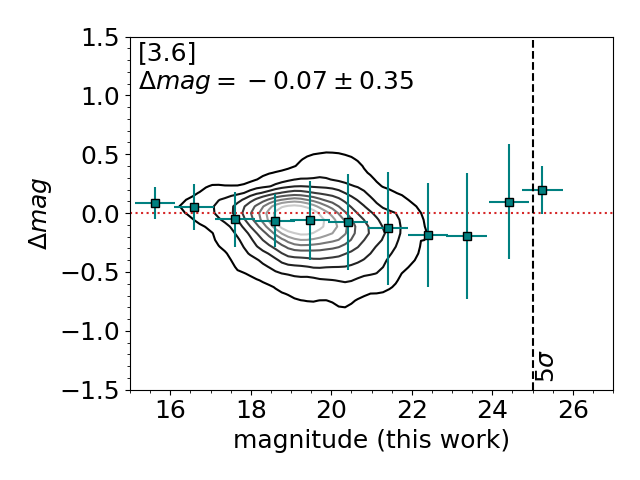}
    \includegraphics[width=\linewidth]{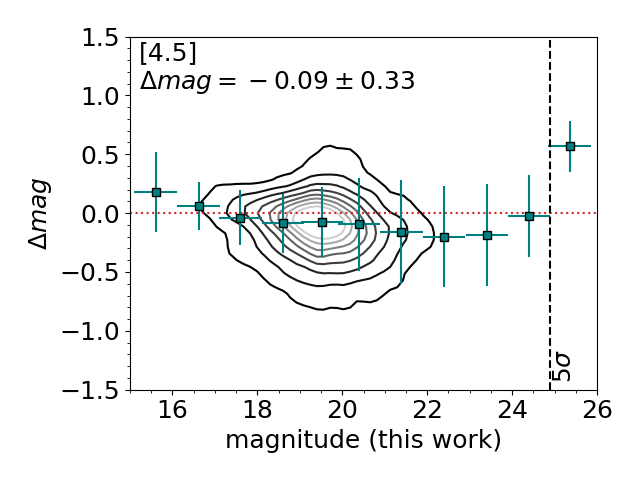}
    \caption{Difference between the IRAC magnitudes derived in this work and the ones included in the Cosmic Dawn Survey catalogue. The black contours show the distribution of all sources in common between the two catalogues (from 10\% to 90\% of the sample), while green squares show the average value in magnitude bins. We report in the top left the mean and standard deviation of the difference in magnitude. The black vertical dashed lines show the magnitude corresponding to the $5\sigma$ depths. }
    \label{fig:IRAC_comp}
\end{figure}

In Figure \ref{fig:IRAC_comp} we show the comparison between the IRAC magnitudes of the sources in common between the two catalogues. Magnitudes are consistent within the two catalogues, with a negative bias below $\Delta mag<0.1$ in both bands, showing that our magnitudes are, on average, slightly brighter.

\section{Catalogue description}\label{sec:cat}
In this Section we describe the columns available in the released catalogue, which includes the optical to radio photometry of LOFAR sources as well as their photometric redshifts. When a column is not available for a specific source, the value is fixed to -99.9. This catalogue is made publicly available\footnote{Add a link to the repository.}, where we also include the update version of the catalogue released by \citet{Bondi2024}. 

\begin{enumerate}
    \item Source\_id: ID of the radio source. 
    \item RA: Original radio radio right ascension [deg]
    \item E\_RA: uncertainties of the right ascension [deg]
    \item RA\_new: radio right ascension after astrometry correction (see Sec. \ref{sec:offsets}) [deg]
    \item DEC: original radio declination [deg]
    \item E\_DEC: uncertainties of the radio declination [deg]
    \item RA\_new: radio declination after astrometry correction (see Sec. \ref{sec:offsets}) [deg]
    \item Total\_flux: total radio flux [Jy]
    \item E\_Total\_flux: uncertainties on the total radio flux [Jy]
    \item Peak\_flux: peak radio flux [Jy/beam]
    \item E\_Peak\_flux: uncertainties on the peak radio flux [Jy/beam]
    \item Maj: major axis[deg]
    \item E\_Maj: uncertainties on the major axis [deg]
    \item Min: minor axis [deg]
    \item E\_Min: uncertainties on the minor axis [deg]
    \item PA: position angle[deg]
    \item E\_PA: uncertainties on the position angle[deg]
    \item S\_Code: classification from \texttt{PyBDSF}
    \item mask: if the radio source is inside the star mask
    \item ID\_flag: 0 if the source is detected both in optical and near-IR. 1 if the source is IRAC-only and 2 if it is optical-only.
    \item ID\_IRAC: ID of the IRAC source 
    \item RA\_IRAC: IRAC right ascension [deg]
    \item DEC\_IRAC: IRAC declination [deg]
    \item MAG\_AUTO\_CH1: Kron AB magnitude in the [3.6$\,\rm\mu m$] filter
    \item MAGERR\_AUTO\_CH1: error of the Kron AB magnitude in the [3.6$\,\rm\mu m$] filter
    \item MAG\_AUTO\_CH2: Kron AB magnitude in the [4.5$\,\rm\mu m$] filter
    \item MAGERR\_AUTO\_CH2: error of the Kron AB magnitude in the [4.5$\,\rm\mu m$] filter
    \item MAG\_AUTO\_CH3: Kron AB magnitude in the [5.8$\,\rm\mu m$] filter
    \item MAGERR\_AUTO\_CH3: error of the Kron AB magnitude in the [5.8$\,\rm\mu m$] filter
    \item MAG\_AUTO\_CH4: Kron AB magnitude in the [8.0$\,\rm\mu m$] filter
    \item MAGERR\_AUTO\_CH4: error of the Kron AB magnitude in the [8.0$\,\rm\mu m$] filter
    \item ID\_opt: ID of the optical source, as taken from \citet{Taylor2023}
    \item RA\_opt: optical right ascension [deg]
    \item DEC\_opt: optical declination [deg]
    \item g\_kron\_mwc: Kron AB magnitude in the $g$ filter, corrected for galactic extinction
    \item g\_kronerr: error of the Kron AB magnitude in the $g$ filter
    \item r\_kron\_mwc:  Kron AB magnitude in the $r$ filter, corrected for galactic extinction
    \item r\_kronerr: error of the Kron AB magnitude in the $r$ filter
    \item i\_kron\_mwc: Kron AB magnitude in the $i$ filter, corrected for galactic extinction
    \item i\_kronerr: error of the Kron AB magnitude in the $i$ filter
    \item z\_kron\_mwc: Kron AB magnitude in the $z$ filter, corrected for galactic extinction
    \item z\_kronerr: error of the Kron AB magnitude in the $z$ filter
    \item y\_kron\_mwc: Kron AB magnitude in the $y$ filter, corrected for galactic extinction
    \item y\_kronerr: error of the Kron AB magnitude in the $y$ filter
    \item n921\_kron\_mwc:  Kron AB magnitude in the NB921 filter, corrected for galactic extinction
    \item n921\_kronerr: error of the Kron AB magnitude in the NB921 filter
    \item n816\_kron\_mwc:  Kron AB magnitude in the NB816 filter, corrected for galactic extinction
    \item n816\_kronerr: error of the Kron AB magnitude in the NB816 filter
    \item mask\_optnir: if the optical or NIR source is inside the star mask
    \item multi-opt: is true if multiple matches were possible between optical and NIR
    \item LR\_mc: colour-magnitude LR. Reliable matches have values larger than the threshold $LR_{th}=0.0079$.
    \item LR\_ID: if the sources has been selected only based on the LR, without visual inspection 
    \item PACS\_70: AB magnitude of the PACS blue filter at $70\mu m$
    \item PACS\_70\_err: AB magnitude error of the PACS blue filter at $70\mu m$
    \item PACS\_100: AB magnitude of the PACS green filter at $100\mu m$
    \item PACS\_100\_err: AB magnitude error of the PACS green filter at $100\mu m$
    \item PACS\_160: AB magnitude of the PACS red filter at $160\mu m$
    \item PACS\_160\_err: AB magnitude error of the PACS red filter at $160\mu m$
    \item SPIRE\_250: AB magnitude of the SPIRE short filter at $250\mu m$
    \item SPIRE\_250\_err: AB magnitude error of the SPIRE short filter at $250\mu m$
    \item SPIRE\_350: AB magnitude of the SPIRE medium filter at $250\mu m$
    \item SPIRE\_350\_err: AB magnitude error of the SPIRE medium filter at $350\mu m$
    \item SPIRE\_500: AB magnitude of the SPIRE long filter at $500\mu m$
    \item SPIRE\_500\_err: AB magnitude error of the SPIRE long filter at $500\mu m$
    \item ID\_DAWN: ID of the source in the Cosmic Dawn Survey catalogue by \citet{Zalesky2024}
    \item RA\_DAWN: right ascension from the Cosmic Dawn Survey [deg]
    \item DEC\_DAWN: declination from the Cosmic Dawn Survey [deg]
    \item CFHT\_u\_mwc: AB magnitude in the $u$ filter, corrected for galactic extinction
    \item CFHT\_u\_err: error of the AB magnitude in the $u$ filter
    \item z\_phot\_bayesian: photometric redshift, is the Bayesian estimation from \texttt{CIGALE}
    \item z\_phot\_bayesian\_err: uncertainties on the photometric redshift, is the Bayesian estimation from \texttt{CIGALE}
    \item z\_phot\_best: photometric redshift, is the estimation from the template with the minimum $\chi^{2}$ from \texttt{CIGALE}
    \item reduced\_chi\_square: reduced $\chi^{2}$ from \texttt{CIGALE}, corresponding to the z\_phot\_best photometric redshift
    \item z\_spec: spectroscopic redshift from DESI
    \item z\_spec\_err: uncertainties on the spectroscopic redshift from DESI
\end{enumerate}

\end{appendix}

\end{document}